\documentclass[5p, twocolumn]{elsarticle}
\usepackage{xparse}                  
\usepackage{etoolbox}               

\usepackage{amssymb}
\usepackage{amsmath}
\usepackage{todonotes}

\journal{Journal of Systems and Software}

\begin{document}

\begin{frontmatter}



\title{A Research Agenda on Agents and Software Engineering: \\Outcomes from the Rio A2SE Seminar}


\author[SDU]{Davide Taibi \corref{cor1}} 
\ead{taibi@imada.sdu.dk}


\author[AQ]{Henry Muccini \corref{cor1}} 
\ead{henry.muccini@univaq.it}

\author[IIIT]{Karthik Vaidhyanathan\corref{cor1}}
\ead{karthik.vaidhyanathan@iiit.ac.in}

\author[PUC]{Marcos Kalinowski\corref{cor1}} 
\ead{kalinowski@inf.puc-rio.br}

\cortext[cor1]{Lead Organizers}

\address[SDU]{University of Southern Denmark, Vejle, Denmark}
\address[AQ]{FrAmeLab@SWEN, University of L'Aquila, Italy}
\address[IIIT]{Software Engineering Research Center, IIIT Hyderabad, India}
\address[PUC]{Pontifical Catholic University of Rio de Janeiro, Rio de Janeiro, Brazil}

\author{\vspace{1em} \newline Michele Albano,
Antonio Pedro Santos Alves,
Renato Cerqueira,
Mateus Devino,
Matteo Esposito,
Rodrigo Falc\~{a}o,
Vinicius Henning,
Foutse Khomh,
Valentina Lenarduzzi,
Qinghua Lu,
Matías Martínez,
Henrique Mello,
Daniel Mendez,
Lucas Romao.
}


\begin{abstract}
The rise of agentic AI is reshaping software engineering in two intertwined directions: agents are increasingly applied to support software engineering tasks, and Agentic AI systems themselves are complex systems that require re-thinking currently established software engineering practices. To chart a coherent research agenda covering the two directions, we organized the A2SE seminar in Rio de Janeiro, bringing together 18 experts from academia and industry. Through structured presentations, collaborative topic clustering, and focused group discussions, participants identified six thematic areas: Governance, Software Engineering for Agents, Agents for Software Architecture, Quality and Evaluation, Sustainability, and Code, and they prioritized short-term and long-term research directions for each. This paper presents the resulting community-driven, opinionated research agenda, offering the SE community a structured foundation for coordinating efforts at this critical juncture.
\end{abstract}


\begin{keyword} Engineering Agentic Systems, Agents for Software Engineering, Research Roadmap



\end{keyword}

\end{frontmatter}

\section{Introduction}
The emergence of agentic AI represents a paradigm shift in software engineering, from models that assist to agents that act autonomously. This shift introduces both opportunities and fundamental challenges for the software engineering research and practice community. On the one hand, agents offer significant promise for automating complex tasks such as architectural decision-making, code generation and repair, requirements analysis, and compliance verification. On the other hand, their non-deterministic behaviour, opacity, and evolving autonomy raise pressing concerns around governance, quality assurance, maintainability, and the appropriate boundaries of human–agent collaboration. These concerns demand a rethink of core software engineering principles, from architectural design to software process models.

Since there are many emerging and promising contributions in the community on using GenAI for software engineering practices and processes, as well as on engineering GenAI-enabled, in particular, Agentic AI systems, this requires consolidation, giving rise to the need for a coherent community-driven research agenda that looks at the existing research landscape and prioritizes the research directions across both short-term and long-term horizons. Prior discussion/workshops/seminar efforts in adjacent areas have demonstrated the value of structured expert deliberation in surfacing cross-cutting challenges that individual research groups may overlook~\cite{lewis2024software,russo2024generative}.

To this end, we organised the A2SE (Agents and Software Engineering) seminar\footnote{https://sites.google.com/view/a2se2026/home} at the Software \& AI Engineering Lab (SAIL) of PUC-Rio, bringing together 18 researchers from academia and industry across five continents, with expertise in Agentic AI spanning different aspects of the software development lifecycle. Through a structured protocol of presentations, collaborative topic clustering, focused group discussions, and qualitative analysis, the seminar produced a research agenda organized around six thematic topics: Governance, Software Engineering for Agents, Agents for Software Architecture, Quality and Evaluation, Sustainability, and Code. For each area, we distinguish between short-term priorities amenable to near-term investigation and long-term challenges requiring sustained community effort.

This paper presents the outcomes of the seminar as an opinionated research agenda with a prioritized set of directions reflecting the collective judgment of the participants. We aim to provide the SE community with a structured starting point for coordinating research efforts at this critical juncture, where the engineering foundations for agentic systems are still being defined while their application to software engineering tasks is rapidly expanding.

\section{Study Design}

\subsection{Seminar Protocol}
We conducted the seminar as focus groups, with experts on Agents from industry and academia as participants. 
The goal of the study was to identify the short- and long-term research agenda for AI Agents in software engineering. 

The seminar followed a protocol combining individual presentations, collaborative clustering, focused group discussions, and qualitative analysis, drawing on established seminar formats used in prior community events~\cite{lewis2024software}.

\subsection{Participants}
18 participants working at the intersection of agentic AI and software engineering were invited to participate in the seminar. 
5 participants were industry practitioners, while the remaining 13 were academics (10 professors, and 3 junior researchers), all with experience in AI Agents and academia-industry collaborations. 

Their skills span the foundational architectural design of large-scale agentic systems, including defining patterns and tackling system reconstruction. A significant focus across the group is on non-functional quality attributes, particularly the sustainability (energy consumption, technical debt, and maintainability) and trustworthiness (reliability, security, testing, and verification) of agents, especially in complex, mission-critical, or regulated environments. Additionally, the researchers address the practical scaling of agent adoption across the software development lifecycle, covering DevOps automation, requirements engineering, knowledge management, and optimizing human-agent collaboration.

\subsubsection{Phase 1: Preparation and Scoping}
Before the seminar, each participant was asked to prepare a five-minute presentation addressing three aspects: (i)~their current work with AI agents, (ii)~what they consider the most important challenges or opportunities in the area, and (iii)~their related publications. This preparatory step ensured that participants arrived with articulated positions, enabling productive discussions from the outset.

\subsubsection{Phase 2: Presentations and Topic Elicitation (Day~1)}
On the first day, each participant delivered their five-minute presentation. Following the presentations, participants were asked to capture their research topics on individual notes and place them on a common board. The organizers, acting as moderators, then guided the participants through an iterative clustering exercise in which topics were grouped into common themes. This process continued until the group reached consensus, resulting in six emergent areas:
i) Governance  ii) Software Engineering for Agents iii) Agents for Software Architecture (Agents4SA); iv) Quality and Evaluation; v) Sustainability and vi) Code.

\subsubsection{Phase 3: Prioritization and Focused Group Discussions (Days~1--2)}
Participants voted on the topics they found most relevant for deeper discussion. Given the group size of sixteen, three parallel working groups were formed, each focusing on one thematic area. Discussions were organized in structured 40-minute sessions, separated by 10-minute breaks, after which topics were rotated between groups to ensure cross-pollination of perspectives. Over the course of two days, a total of six 40-minute focused discussion sessions were conducted.

During the discussions, it was observed that Human-in-the-Loop had substantial overlap with Governance, and the participants collectively decided to merge these into a single area. On the second day, the number of parallel groups was reduced from three to two, as the remaining topics were oriented toward longer-term research challenges requiring deeper deliberation. Throughout the discussions, participants used post-it notes to capture and distinguish short-term and long-term research directions for each thematic area.

\subsubsection{Phase 4: Data Consolidation and Analysis (Day~3)}
On the third day, following the conclusion of the seminar, the four organizers reviewed the post-it notes collected from all groups and transcribed the content into a shared spreadsheet. The organizers then performed open coding following the guidelines provided by~\cite{ref_coding_guidelines}. To support the coding process, ChatGPT was used as an assistive tool for initial code generation, with the organizers reviewing, refining, and validating all codes manually. After four rounds of refinement, we  produced a structured set of research challenges and directions organized by thematic area and time horizon.

\subsubsection{Phase 5: Validation and Feedback from the Participants}
The results of the data analysis were sent to the participants for validation and to complement any missing information. Their feedback was incorporated, and all participants reviewed and approved the final version of the manuscript.


\section{Study Results}
The participants highlighted six high-level research topics that should be addressed. The short- and long-term research directions for the topics are reported in Table~\ref{table}. 

\begin{table*}[t]
\centering
\small
\begin{tabular}{p{3cm}| p{6.5cm} p{6.5cm}}
\hline
 & \multicolumn{2}{c}{\textbf{Research directions}} \\
\textbf{Topic} & \textbf{Short Terms} & \textbf{Long Terms} \\
\hline

\textbf{Governance} &
Agent Governance Models and Responsibility, Policy-Aware and Compliant Agents, Quality Attributes for Governable AI, AgentOps
&
Certification of Agentic Systems, Multi-Layer Governance Frameworks, Policy Enforcement and Threshold Validation, Human--AI Collaboration Models, Governance Debt \\

\textbf{Software Engineering  for Agents} &
Hybrid Agent-Based Architectures, Multi-Agent System Design and Granularity, Agent Orchestration and Coordination, Agent Engineering Frameworks, Requirements Engineering for Agents
&
Agentic Design Patterns and Repositories, Technology-Agnostic Engineering Guidelines, Cost-Aware Agent Adoption Frameworks, Community-Driven Knowledge Consolidation, Formal Verification for Agent Systems, Architectural Practices for Non-Deterministic Systems \\

\textbf{Agents for Software Architecture} &
Specification-Driven Architecting, Integrating human decision-making process with Agent Decisions, Development of Benchmarks for evaluating the capabilities of Agents on Software Architecture Tasks, Agentic AI for architectural processes
&
New Maintainability Paradigms, Architectural Education and Upskilling \\

\textbf{Quality and Evaluation} &
Reliability and Robustness Strategies, Bias and Security in Agent Systems
&
Quality Metrics for Agents, Robustness of Agent Ecosystems, Verification of AI-Generated Artifacts, Human-Readable vs Machine-Readable Code, Future of Coding Practices \\

\textbf{Sustainability} &
Resource-Aware Agent Design, Sustainability Metrics and KPIs, Cost and ROI Models for Agents, Sustainability Debt, Sustainability Awareness
&
Standardized and Hardware-Agnostic Benchmarking, Software modernization to agent-oriented green architectures \\

\textbf{Code} &
Code Generation Quality and Integration, Technical Debt in Generated Code, Architecture Modernization via Agents, Reverse Engineering with Agents
&
Agent--System Communication Protocols, Cost Optimization in Code Generation, Maintainability of Generated Code, Runtime Code Generation, Customization of Coding Styles, Decision Frameworks for Agent vs Traditional Development \\

\hline
\end{tabular}
\caption{Research topics with short-term and long-term directions}
\label{table}
\end{table*}

\subsection{Research Agenda}

Participants considered that research on six topics for agentic systems should prioritize different short and long terms research directions. Below we describe the outcomes of the discussions on the six topics.

\subsubsection{Governance}

Participants considered that short-term research on governance for agentic systems should prioritize establishing clear foundations that enable safe, accountable deployment in practice. In the long term, they emphasized the need to advance governance for agentic systems toward scalable, institutionalized, and societally embedded frameworks capable of sustaining increasingly autonomous and complex software systems.

\textit{Short term.} \textbf{Agent Governance Models and Responsibility} concerns the need to define roles, accountability, and liability in agentic systems, including well-defined human--AI responsibility boundaries. \textbf{Policy-Aware and Compliant Agents} encompasses bidirectional interaction between agents and policies, as well as mechanisms to ensure regulatory compliance, requiring machine-readable policies, to name one implication. \textbf{Quality Attributes for Governable AI} addresses the definition and operationalization of reliability, security, transparency, and explainability, enabling systematic assessment and enforcement. Finally, \textbf{AgentOps} concerns the adaptation of software processes and DevOps practices to support the development, deployment, and monitoring of agentic systems. 

\textit{Long term.} \textbf{Certification of Agentic Systems} concerns the development of standards and certification mechanisms, especially for safety-critical domains. \textbf{Multi-Layer Governance Frameworks} encompasses governance across technical, organizational, and societal layers, including agent-generated systems. \textbf{Policy Enforcement and Threshold Validation} addresses mechanisms to ensure systems operate within acceptable boundaries and constraints. \textbf{Human--AI Collaboration Models} concerns the development of patterns, cost models, and competence frameworks for effective collaboration. Finally, \textbf{Governance Debt} captures the conceptualization and measurement of accumulated governance shortcomings over time.

\subsubsection{Software Engineering for Agents (SE4A)}

Participants considered that short-term research in software engineering for agents should focus on establishing foundational architectural and engineering practices for designing, orchestrating, and building agent-based systems effectively. In the long term, they emphasized the need for systematic knowledge consolidation, including the identification of patterns and the adoption of technology-agnostic methodologies to guide the sustainable development and adoption of agent-based systems.

\textit{Short term.} \textbf{Hybrid Agent-Based Architectures} concerns the design of systems combining agent and non-agent components, including distinctions between agent-native and agent-empowered systems and decisions on when to adopt agents. \textbf{Multi-Agent System Design and Granularity} addresses defining agent granularity and selecting appropriate memory, tools, and context at both agent and system levels, including mechanisms to control autonomy. \textbf{Agent Orchestration and Coordination} concerns the limitations of current orchestration approaches, the need for new architectural solutions, and the role of orchestration and choreography in coordinating agents. \textbf{Agent Engineering Frameworks} addresses the need for systematic and standardized approaches to efficiently build agents and rapidly develop proofs of concept. Finally, \textbf{Requirements Engineering for Agents} concerns identifying relevant quality attributes and understanding how requirements engineering differs from traditional approaches.

\textit{Long term.} \textbf{Agentic Design Patterns and Repositories} concerns the creation of repositories of design patterns for agent-based systems, supported by empirical studies to assess their value. \textbf{Technology-Agnostic Engineering Guidelines} addresses the need for general guidelines and workflows that are not tied to specific technologies and remain sustainable over time. \textbf{Cost-Aware Agent Adoption Frameworks} concerns understanding when to use agents, evaluating costs, and supporting trade-off decisions. \textbf{Community-Driven Knowledge Consolidation} addresses the need to organize community initiatives to share practices, align terminology, and exchange experience between academia and industry. Finally, \textbf{Formal Verification for Agent Systems} and \textbf{Architectural Practices for Non-Deterministic Systems} concern the development of methods to ensure correctness and to design systems that account for non-deterministic behavior.

\subsubsection{Agents for Software Architecture (A4SA)}

Participants considered that short-term research on agents for software architecture should focus on integrating agents into architectural processes and on supporting decision-making through structured, measurable approaches. For the long term, they emphasized the need to rethink core software engineering concepts, particularly maintainability and education, in light of the shift toward AI-driven and specification-based systems.

\textit{Short term.} \textbf{Specification-Driven Architecting} concerns the shift from code-centric to specification-driven systems, including the use of ADL-based generation. \textbf{Integrating Human Decision Making Process with Agent Decisions} addresses the interaction between human decision-making and agent-based support in architectural processes. \textbf{Development of Benchmarks for Evaluating the Capabilities of Agents on Software Architecture Tasks} concerns the need for benchmarks to systematically assess agent performance in architectural tasks. Finally, \textbf{Agentic AI for Architectural Processes} addresses the use of agents for knowledge management, analysis, and decision making in software architecture.

\textit{Long term.} \textbf{New Maintainability Paradigms} concerns redefining maintainability when code is no longer the primary artifact. Finally, \textbf{Architectural Education and Upskilling} addresses the need to redesign education and training to embed architectural thinking for AI-driven systems.

\subsection{Quality and Evaluation}

Participants considered that short-term research on quality and evaluation should focus on improving the reliability and trustworthiness of agent-based systems, particularly in terms of robustness, security, and bias mitigation. For the long term, they emphasized the challenges posed by malicious agents and the need to understand and overcoming both agent behavior and generated artifacts, including new quality metrics, verification approaches, and the evolution of coding practices.

\textit{Short term.} \textbf{Reliability and Robustness Strategies} concerns strategies to increase effectiveness, including real-time feedback loops to improve reliability. \textbf{Bias and Security in Agent Systems} addresses dealing with malicious agents and understanding and overcoming biases in foundation models.

\textit{Long term.} \textbf{Quality Metrics for Agents} concerns the definition of new metrics to evaluate agent performance, such as the number of iterations required to achieve acceptable results. \textbf{Robustness of Agent Ecosystems} addresses maintaining robustness despite changes in underlying non-deterministic components and ensuring control boundaries as agents may generate other agents. \textbf{Verification of AI-Generated Artifacts} concerns the increasing role of formal methods and techniques to ensure correctness of generated code. \textbf{Human-Readable vs Machine-Readable Code} addresses the need for generated code to remain interpretable by humans or supported by specialized tools when not human-readable. Finally, \textbf{Future of Coding Practices} concerns the evolution of development practices, including shifts toward specification-driven approaches, reduced emphasis on manual coding, and changes in maintainability and engineering roles.

\subsection{Sustainability}
Participants considered that short-term research on sustainable LLM agents should prioritize laying concrete foundations that enable the design, assessment, and adoption of agents, with explicit attention to resource efficiency, costs, and long-term sustainability impacts in practice. In the long term, they emphasized the need to advance research on sustainable LLM agents toward more general, transferable, and systemic approaches capable of supporting heterogeneous infrastructures and the large-scale modernization of software systems toward greener agentic architectures.

\textit{Short term.} \textbf{Resource-Aware Agent Design} concerns the optimization of model size, token usage, and energy consumption in agentic systems, fostering design decisions that explicitly account for computational efficiency. \textbf{Sustainability Metrics and KPIs} encompasses the definition of measurable indicators to assess the environmental and economic impact of agents, enabling systematic evaluation and comparison. \textbf{Cost and ROI Models for Agents} addresses the need to evaluate the financial viability of agentic solutions, including both human-related and computational costs. \textbf{Sustainability Debt} concerns the conceptualization and tracking of the long-term sustainability implications that accumulate over time through agent use. Finally, \textbf{Sustainability Awareness} addresses mechanisms to inform users about the sustainability costs associated with employing agents, thereby supporting more informed and responsible use.

\textit{Long term.} \textbf{Standardized and Hardware-Agnostic Benchmarking} concerns the definition of benchmarking techniques for agentic systems that remain independent of specific hardware infrastructures, thus enabling broader comparability and reproducibility. Finally, \textbf{Software Modernization to Agent-Oriented Green Architectures} addresses the migration of complex legacy applications to agentic systems designed according to sustainability principles, enabling greener software architectures at scale.

\subsection{Code}
Participants considered that short-term research on code and development should focus on ensuring that code generation can be reliably adopted in practice, particularly in terms of quality, integration, and modernization of existing systems. In the long term, they emphasized the need to understand how agent-based and code-generation approaches can be integrated into software systems, including communication, maintainability, and decision-making between agent-based and traditional development approaches.

\textit{Short term.} \textbf{Code Generation Quality and Integration} concerns ensuring the quality of generated code and its effective integration with legacy systems. \textbf{Technical Debt in Generated Code} addresses the challenges related to the technical debt introduced by generated code, including its accumulation over time. \textbf{Architecture Modernization via Agents} concerns migrating across languages and modernizing architectures to improve specific qualities, such as reducing coupling. Finally, \textbf{Reverse Engineering with Agents} addresses how to perform reverse engineering at runtime and statically, including whether LLMs and agents can outperform traditional static analysis.

\textit{Long term.} \textbf{Agent–System Communication Protocols} concerns how agents and systems should communicate, including whether to rely on MCP, other protocols, or a mix of protocols. \textbf{Cost Optimization in Code Generation} addresses reducing the cost of code generation, for example, by spending fewer tokens. \textbf{Maintainability of Generated Code} concerns the extent to which the code generated by agents is maintainable and whether code generated by agents remains maintainable after multiple generations. \textbf{Runtime Code Generation} refers to the ability of systems to self-code at runtime, i.e., to adapt to new requirements by changing their source code autonomously, compiling and deploying it on-the-fly. Concerns about reliability, scope, and context are expected to play a major role in this area. \textbf{Customization of Coding Styles} addresses methods to adapt generated code to company practices. Finally, \textbf{Decision Frameworks for Agent vs Traditional Development} concerns understanding when to develop a feature as an agent or using traditional approaches.

\section{Discussion}

The results reveal several cross-cutting themes that highlight key challenges for agent-based systems.

\textbf{Engineering foundations} across \textbf{SE4Agents} and \textbf{Agents4SA} show the need for new architectural models, orchestration mechanisms, and development frameworks to handle system complexity and non-determinism.

\textbf{Governance, accountability, and policy integration} emphasize that responsibility, controllability, and compliance must be embedded into system design rather than treated as external concerns.

\textbf{Standardization and pattern reuse} reflect the need to consolidate emerging practices into reusable patterns, guidelines, and shared knowledge.

\textbf{Human--AI collaboration and skills evolution} point to significant socio-technical changes, including new roles, competencies, and decision-making processes.

Finally, \textbf{quality, evaluation, and sustainability} highlight the need for new metrics, evaluation approaches, and explicit consideration of long-term trade-offs.

Further, we postulate that since the themes are cross-cutting but also overlapping (e.g. governance and policy integration requires, same as others, standardisation), we cannot treat them in isolation. Rather, they demonstrate that agentic systems require a holistic approach, where technical, organizational, and human aspects are tightly interconnected.

\section{Conclusions}
We presented a community-driven research agenda on agents and software engineering, identifying short-term priorities for practical adoption and long-term challenges requiring fundamental changes in architecture, quality, governance, and development practices. Participants highlighted the impact of non-determinism, the shift toward specification-driven development, and the integration of agents into existing systems.

The seminar also underscored the importance of continued community efforts, such as focused seminars and industry events, to consolidate knowledge, align terminology, and bridge academia and practice. Additionally, adopting agent-based systems entails broader process and organizational changes, including new skills and evolving roles for engineers.

Last, participants highlighted an important possible direction of Agents for Science, including agents for conducting empirical studies such as thematic coding, or literature reviews that call for a reflection in the light of a balance of the advantages in terms of efficiency and the risks that decoupling researchers from data can bring to the researchers' training process.

Finally, the results emphasize the need for continuous empirical validation and knowledge consolidation to support the sustainable evolution of agent-based software engineering.

\section*{Acknowledgments}

The authors used generative AI tools to assist in writing this paper. Grammarly was used for spelling, grammar, and readability improvements. All technical ideas, analyses, results, and conclusions in this paper were conceived, developed, and verified by the authors. The authors take full responsibility for the content of the final manuscript.\\

\bibliographystyle{elsarticle-num}
\bibliography{main}
\end{document}